\input{aipcheck}

\documentclass[
    ,final            
  ]
  {aipproc}

\layoutstyle{6x9}

\begin{document}

\title{Matter content in AGN jets:
 constraint from cocoon dynamics?}


\author{M. Kino}{
  address={SISSA, via Beirut 2-4, 34014 Trieste, Italy}
}

\author{N. Kawakatu}
{
  address={SISSA, via Beirut 2-4, 34014 Trieste, Italy}
}

\begin{abstract}

The matter content of jets in active galactic nuclei
is examined in a new way.
We model the dynamical expansion of its cocoon
embedded in the intra-cluster medium (ICM).
By comparing the observed shape of the cocoon with
that expected from the theoretical model,
we estimate the total pressure ($P_{\rm c}$) and 
electron temperature ($T_{e}$) of the cocoon.
The number density of the total electrons ($n_{e^{-}}$)
is constrained 
by using the non-thermal spectrum  of the hot spot
and
the analysis of the momentum balance 
between the jet thrust and the rum pressure of ICM.
Together with the obtained $P_{\rm c}$,  $T_{e}$ 
and $n_{e^{-}}$,
we constrain the matter content in the jets.
We find that, in the case of Cygnus A, 
the ratio of number density of protons to that
of electrons  
is of order of $10^{-3}$ .
This implies the existence of a large number of positron 
in the jet.

\end{abstract}

\maketitle


\section{Introduction}

\begin{figure}\label{fig:CygA}
 \includegraphics[width=.33\textheight]{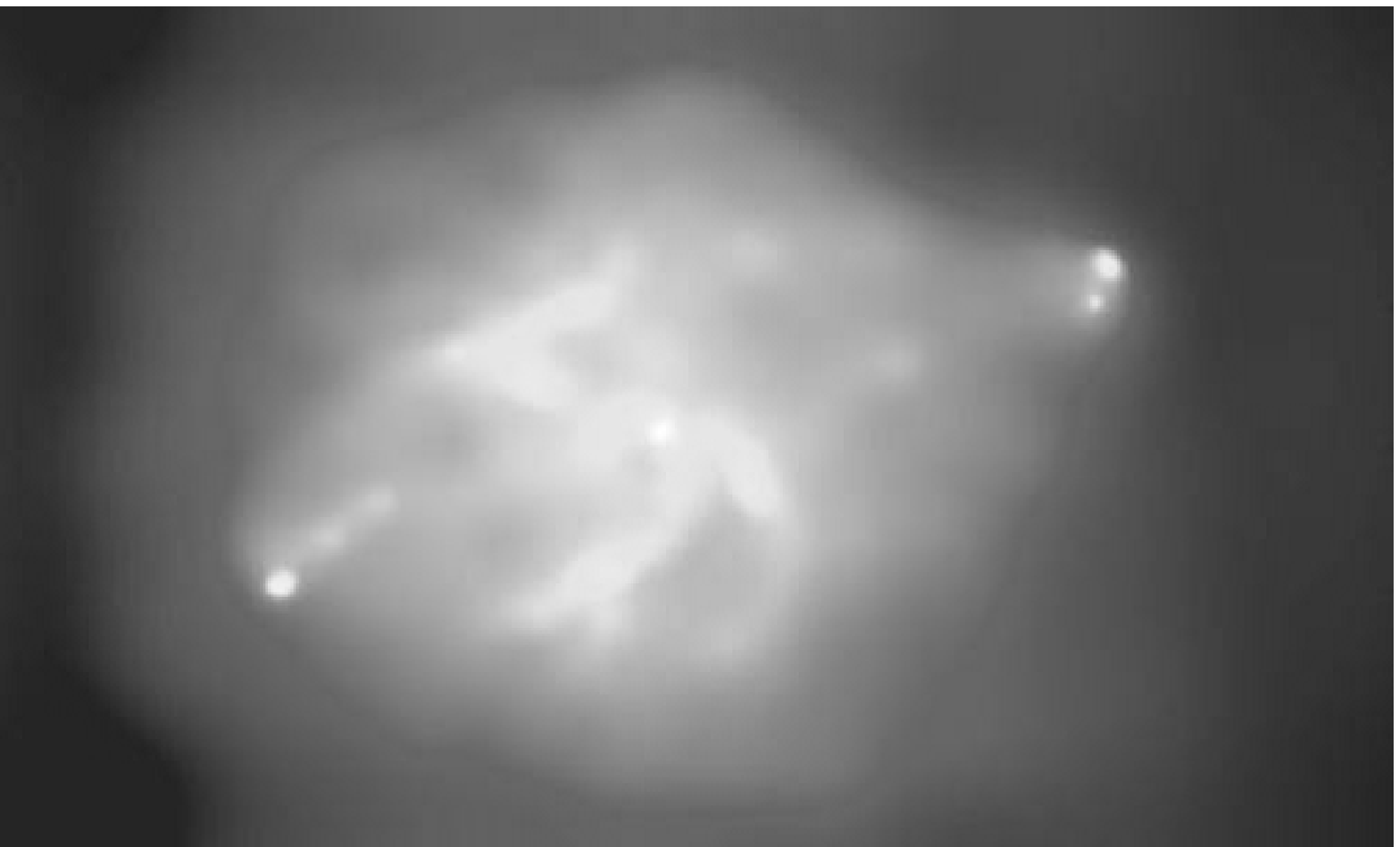}
\hspace*{0.3cm}
\mbox{\raisebox{-1cm} {\includegraphics[height=.33\textheight]{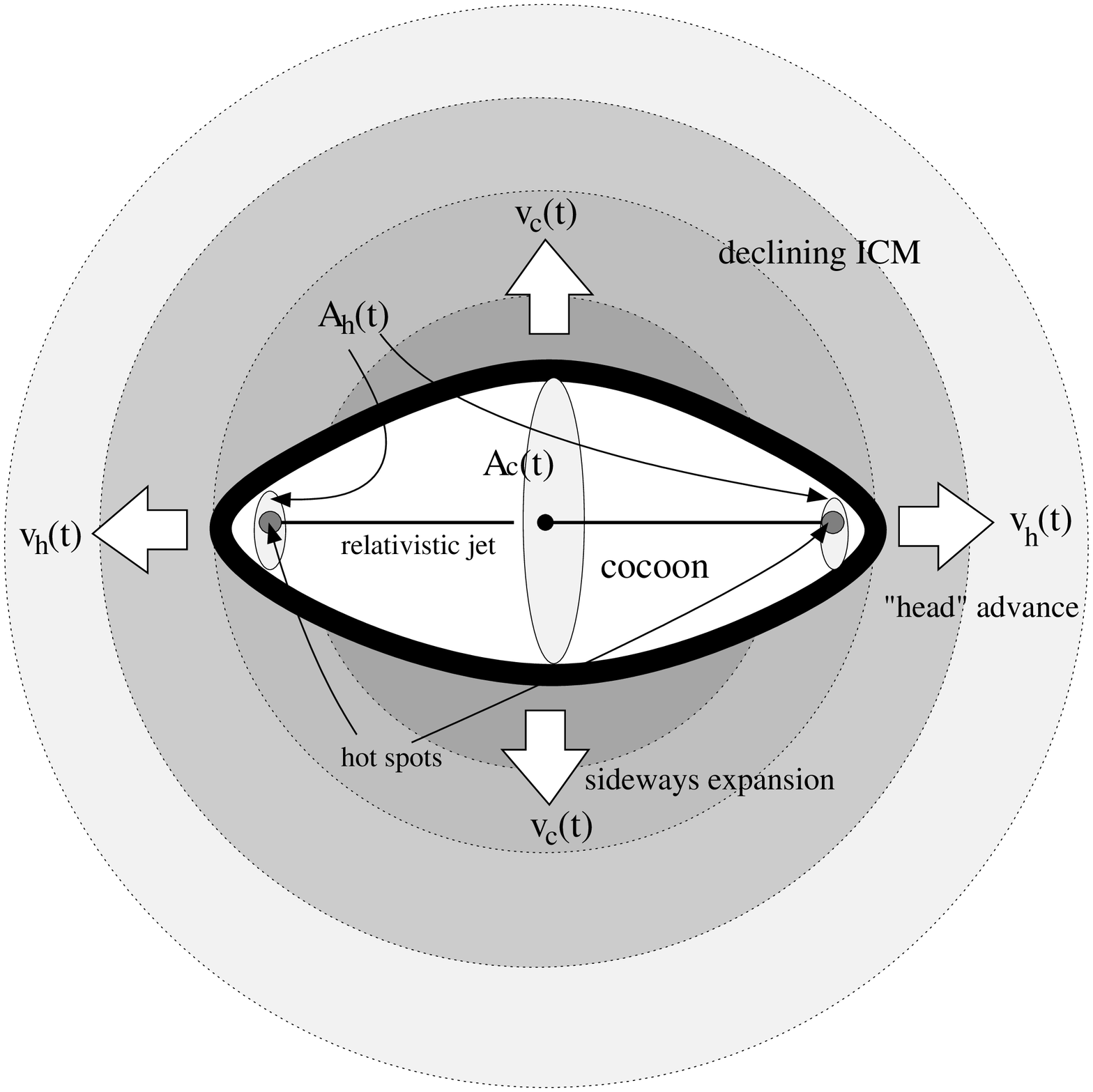}}}
  \caption{Left: X-ray image of Cygnus A by {\it Chandra} \cite{W00}. 
Right: A cartoon of the cocoon expansion in powerful FR II 
source \cite{BC89,KK05}.}
\end{figure}

The matter content in extragalactic jet 
is one of the 
primal issues for resolving the jet formation
mechanism in active galactic nuclei (AGNs) \cite{BBR}.
However it has been a longstanding problem over the years, 
since it is hard to get the electromagnetic signal 
from the component
such as {\it thermal electrons and/or protons} 
co-existing with non-thermal electrons.
So far, at the sub-pc scale inner jet,
mainly three approach have been proposed
to constrain the plasma content in AGN jets.
They are
based on the (i) synchrotron self-absorption analysis
\cite{CF93,R96,h99},
(ii) the observed circular polarization 
\cite{W98}, and  
(iii) the constraint from the absence of 
bulk-Compton emission \cite{SM00}.
As a complementary approach,  
the constraints from the hydrodynamical interaction between
the  large scale jet (100kpc-Mpc) and intra-cluster medium (ICM)
has been recently proposed \cite{KT04}.
In the proceeding,
a new approach to constrain on the matter content is presented, 
by the combination of cocoon dynamics 
and non-thermal emission analysis at the hot spot (see Fig. \ref{fig:CygA}).

\section{the cocoon model}

Here we define a key
parameter $\eta$ describing the degree of baryon loading
\begin{eqnarray} \label{eq:eta}
n_{p}\equiv
\eta n_{e^{-}}  ,
\end{eqnarray}
where $n_{p}$ and 
$n_{e^{-}}$ are
the total number densities of 
protons and 
electrons in the cocoon, respectively.
The case of $\eta=0$ corresponds to pure $e^{\pm}$ plasma
while  $\eta=1$ corresponds to the pure electron-proton plasma.
The total number densities of 
positron is expressed as
$n_{e^{+}}=(1-\eta)n_{e^{-}}$ by
the charge neutral condition.
The number density of 
non-thermal (hereafter ``NT'') electrons  
is written as $n_{e}^{\rm NT}$. 
Main assumptions in this work are as follows;
(i)  a jet velocity $v_{\rm j}(=\beta_{\rm j}c)$
is a relativistic one,
(ii) the kinetic power $L_{\rm j}$ and mass flux $J_{\rm j}$
of the jet is constant in time,
(iii) $L_{\rm j}$ is
all deposited into the cocoon as the internal energy \cite{S74},
(iv) the shocked matter pressure
is dominated by thermal plasma, 
and 
(v) the electrons and protons are separately
thermalized by shock heating and they become two temparature phase
(i.e., $(T_{e}=\frac{m_{e}}{m_{p}}T_{p})$
where $T_{e}$ and $T_{p}$ are
the electron (and positron) temperature
and proton temperature).

The time-averaged 
mass and energy injection from the relativistic jet into
the cocoon, which govern 
pressure $P_{\rm c}$ and 
mass density $\rho_{\rm c}$ of the cocoon  
are written as
\begin{eqnarray}\label{eq:pc}
\frac{1}{{\hat \gamma}_{c}-1}
\frac{P_{\rm c}V_{c}}{t_{\rm age}}=
T_{\rm j}^{01} 
A_{\rm j}, \qquad
\frac{\rho_{\rm c}V_{c}}{t_{\rm age} }=
J_{\rm j} 
A_{\rm j}   ,
\end{eqnarray}
where
${\hat \gamma}_{c}$,
$V_{\rm c}$,
$t_{\rm age}$,
$T_{\rm j}^{01}$, 
$J_{\rm j}$,
$A_{\rm j}$,
are
the adiabatic index of the plasma in the cocoon,
the volume of the cocoon,
the source age,
the energy and mass flux of the jet, and 
the cross-sectional area of the jet,
respectively.
The condition of cold jet lead to 
$T_{\rm j}^{01}=\rho_{\rm j}c^{2}\Gamma_{\rm j}^{2}v_{\rm j}$,
$J_{\rm j}=\rho_{\rm j}\Gamma_{\rm j}v_{\rm j}$ where
$\rho_{\rm j}$, and 
$\Gamma_{\rm j}$ are
mass density and 
bulk Lorentz factor of the jet \cite{BR74}.
It is useful to define the ratio of 
"the volume swept by the unshocked relativistic jet"
to "the volume of cocoon" which is written as 
${\cal A}\equiv
(A_{\rm j}v_{\rm j} t_{\rm age})/V_{\rm c}$.
Here we set 
$A_{\rm j}=\pi R_{\rm hs}^{2}$ and 
$V_{\rm c}=(2\pi/3){\cal R}^{2}l_{\rm hs}^{3}$,
where 
$R_{\rm hs}$, 
${\cal R}$, and 
$l_{\rm hs}$ are
the size of the hot spot,
the aspect-ratio of the cocoon and 
the distance from the central core to
the hot spot, respectively.
Together with these basic equations and 
two temperature condition,
we can express  
$T_{e}$, 
$T_{p}$, 
$n_{e}$, and 
$n_{p}$ as
\begin{eqnarray}\label{eq:Te}
3 \Gamma_{j}c^{2} = \frac{kT_{e}}{m_{e}} =
 \frac{kT_{p}}{m_{p}}, \quad
n_{e^{-}}=
\frac{\Gamma_{j}\rho_{j}{\cal A}}
{(2-\eta)m_{e}+\eta m_{p}}, \quad
n_{p}= \eta n_{e^{-}} .
\end{eqnarray}

\section{constraints on the baryon loading}

In principle, we can determine $\eta$
if we know the value of 
$T_{e}$,
$n_{e^{-}}$, and 
$P_{\rm c}$, 
since the sum of the pressures of electrons, positrons and protons 
is written as 
$P_{\rm c}=(n_{e^{-}}+n_{e^{+}}) k T_{e}
+n_{p} k T_{p}=
(2-\eta)n_{e^{-}} kT_{e}+\eta n_{e^{-}} (m_{p}/m_{e})kT_{e}$.
Regarding $k T_{e}$, we have directly obtained the result of 
$k T_{e}=3\Gamma_{\rm j}m_{e}c^{2}$ (in Eq. (\ref{eq:Te})). 
As for $P_{\rm c}$, we independently obtained it 
by solving the dynamics of cocoon expansion \cite{KK05}.
$n_{e^{-}}$ can be constrained  
with the aid of the observational property of
the hot spot (see details in \cite{KT04,KK05}).
In general, the number density
of NT electron is smaller than that of total particles. 
From this, we can estimate
the minimum value of the total electron as 
$n_{e^{-},\rm min} =
min[\Gamma_{\rm j}n_{\rm j,e^{-}}{\cal A}]  
\simeq 5 \times 10^{-5}       
(n_{e}^{\rm NT}/10^{-3}{\rm cm^{-3}})
({\cal A}/0.05)$
where $n_{e}^{\rm NT}$
is the number density of NT electrons in the hot spot.
Here we used the relativistic shock junction between the 
jet and hot spot \cite{BM76}.
The upper limit of 
the number density of total electrons 
can be obtained
by solving the balance between
the thrust of the jet and the ram pressure of 
the ICM.
The maximum $n_{e^{-}}$ corresponds to
the case of 
pure $e^{\pm}$ plasma sustaining the 
ram pressure of the ICM.
$n_{e^{-},max}=
max[\Gamma_{\rm j}\rho_{\rm j}{\cal A}/m_{e}]
\simeq 1 \times 10^{-3}
(\Gamma_{\rm j}/10)
({\cal A}/0.05)
(n_{ICM}/10^{-2}{\rm cm^{-3}})$.
Here the mass-density ratio can be estimated as 
$\rho_{\rm j,max}/\rho_{\rm ICM}
=(\beta_{hs}/\Gamma_{j})^{2}\sim 10^{-4}$
(Eq. (8) in \cite{KT04}).

\begin{figure}\label{fig:np}
 \includegraphics[height=.35\textheight]{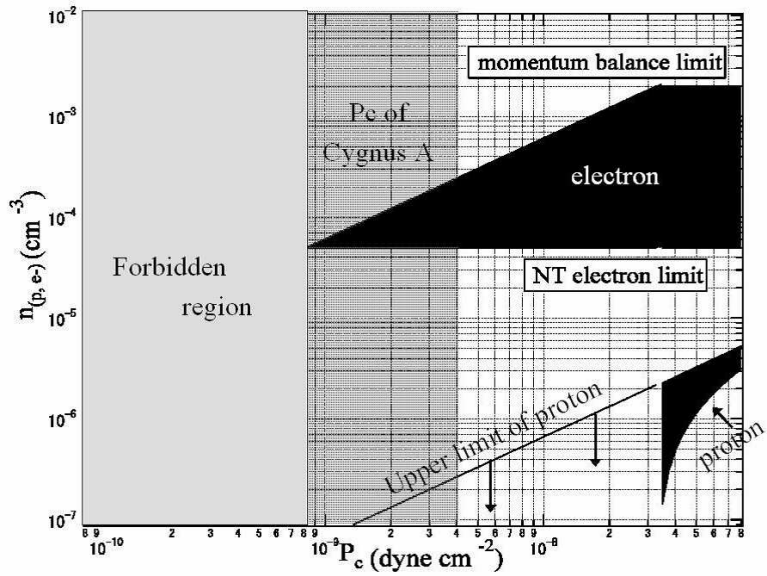}
  \caption{The number densities of electrons and protons in Cygnus A.
The ICM pressure is estimated as $P_{\rm a}= 8\times 10^{-11}{\rm dyne \ cm^{-2}}$ \cite{A84}.
The proton number density is order of $\sim 10^{-3}$ of electrons.
}
\end{figure}

In Fig. \ref{fig:np},
we show the resultant $n_{e^{-}}$ and
$n_{p}$ of Cygnus A.
Based our study of Cygnus A \cite{KK05,KK05b},
we estimate the allowed range of $P_{\rm c}$ as
$8\times 10^{-10}{\rm dyn \ cm^{-2}}
<P_{\rm c}<4\times 10^{-9}{\rm dyn \ cm^{-2}}$
is $P_{\rm c}$ (in gray-color) which is consistent with 
other estimate of \cite{BC89,C98}. 
Since we do not have a consistent value of 
$n_{e^{-}}$
below $P_{c}=8\times 10^{-10}{\rm dyn \ cm^{-3}}$,
the region  below it is ruled out.
It is useful to define the critical cocoon pressure
as $P_{\rm c}^{*}\equiv max[2n_{e^{-}}kT_{e}]=3\times 10^{-8}
{\rm dyn \ cm^{-2}}$.
When $P_{c}>P_{\rm c}^{*}$ is satisfied,
the baryon loading  $\eta$ is determined by
$\eta=(2m_{e}/m_{p})(P_{c}-P_{\rm c}^{*})/P_{\rm c}^{*}$.
In other words, the baryon loading is inevitably required
to support the cocoon.
Furthermore, because of the condition of   
$ n_{p} k T_{p} > (n_{e^{-}}+n_{e^{+}}) k T_{e}$,
 ``proton-supported'' cocoon will be realized.
Below $P_{\rm c}^{*}$, 
$\eta$ only has an upper limit such as  
$0\le \eta<(2m_{e}/m_{p})(P_{c}-P_{\rm c}^{*})/P_{\rm c}^{*}$.
We find that 
the Cygnus A corresponds to this regime and 
the baryon loading is 
$\eta\sim {\cal O}(10^{-3})$.

\section{Summary and discussion}

The matter content of jets in active galactic nuclei
is examined.
The key point is that the quantities of 
$T_{e}$,
$n_{e^{-}}$, and 
$P_{\rm c}$ can be constrained independently.
By comparing  theoretical model of cocoon expansion
and observation, we constrain the $P_{\rm c}$ and  $T_{e}$.
The value of $n_{e^{-}}$ is constrained
is constrained 
by using the non-thermal spectrum  of the hot spot
and
the analysis of the momentum balance 
between the jet thrust and the rum pressure of ICM.
Combining these quantities,
we constrain the matter content in the jets.
The analysis is focused on Cygnus A.
We find that 
$\eta\sim {\cal O}(10^{-3})$ in the case 
of Cygnus A.
As for the number density,
this agree with 
the suggestion of larger number density of $e^{-}e^{+}$ plasma
by \cite{KT04, SM00,S05}.
Furthermore \cite{SM00,S05} 
suggest that kinetic power of 
baryon is larger than that of leptons. 
We keep this as the important future investigation.


{\it Acknowledgements:}
We thank A. Celotti, L. Stawarz and
A. Mizuta for valuable comments.
We acknowledge the Italian MIUR and
INAF financial supports.



\bibliographystyle{aipproc}   

\bibliography{sample}

\IfFileExists{\jobname.bbl}{}
 {\typeout{}
  \typeout{******************************************}
  \typeout{** Please run "bibtex \jobname" to optain}
  \typeout{** the bibliography and then re-run LaTeX}
  \typeout{** twice to fix the references!}
  \typeout{******************************************}
  \typeout{}
 }

\end{document}